# Exploration of new 212 MAB phases: M$_2$AB$_2$ (M=Mo, Ta; A=Ga, Ge) via DFT calculations

A. K. M Naim Ishtiaq[1,2#], Md Nasir Uddin[1,2#], Md. Rasel Rana[1,2], Shariful Islam[1,2], Noor Afsary[1], Karimul Hoque[1,*], Md. Ashraf Ali[2,†]

[1] Physics Discipline, Khulna University, Khulna 9208, Bangladesh.

[2] Advanced Computational Materials Research Laboratory (ACMRL), Department of Physics, Chittagong University of Engineering and Technology (Chattogram-4349, Bangladesh.

*hoquekarimul@phy.ku.ac.bd, †ashrafphy31@cuet.ac.bd

## Abstract

The recently developed MAB phases, an extension of the MAX phase, have sparked interest in research among scientists because of their better thermo-mechanical properties. In this paper, we have explored four new MAB phases M$_2$AB$_2$ (M=Mo, Ta and A=Ga, Ge) and studied the elastic, electronic, thermal, and optical properties to predict the possible applications. The stability of the new phases has been confirmed by calculating formation energy (E$_f$), formation enthalpy ($\Delta H$), phonon dispersion curve (PDC), and elastic constant (C$_{ij}$). The study reveals that M$_2$AB$_2$ (M=Mo, Ta and A=Ga, Ge) exhibit significantly higher elastic constants, elastic moduli, and Vickers hardness values than their counterpart 211 borides. Higher Vickers hardness values of Ta$_2$AB$_2$ (A=Ga, Ge) than Mo$_2$AB$_2$ (A=Ga, Ge) have been explained based on the values of the bond overlap population. The analysis of the density of states and electronic band structure revealed the metallic nature of the borides under examination. The thermodynamic characteristics of M$_2$AB$_2$ (M=Mo, Ta and A=Ga, Ge) under high temperatures (0–1000 K) are investigated using the quasi-harmonic Debye model. Critical thermal properties such as melting temperature (T$_m$), Grüneisen parameter ($\gamma$), minimum thermal conductivity (K$_{min}$), Debye temperature ($\Theta_D$), and others are also computed. Compared with 211 MAX phases, the 212 phases exhibit higher values of ($\Theta_D$) and T$_m$, along with a lower value of K$_{min}$. These findings suggest that the studied compounds exhibit superior thermal properties that are suitable for practical applications. The optical characteristics have been examined, and the reflectance spectrum indicates that the materials have the potential to mitigate solar heating across various energy regions.

#First two authors contributed equally.

1. Introduction

The MAX phase has garnered significant attention in the present era due to its outstanding mechanical and thermal characteristics at high temperatures, showcasing attributes shared by both metals and ceramics. The increased interest in MAX phase materials can be traced back to Barsoum's noteworthy contributions [1], [2]. The term MAX phase represents a family of multilayer solids where M is an earlier transition metal, A is an element from the IIIA or IVA group of the periodic table, and X is an atom of C/N/B [3]. MAX phase materials showcase metallic behavior due to alternate metallic A-layers and ceramic behavior attributed to the MX layers [4]. Like most metals and alloys, MAX phase materials have excellent thermal shock resistance, superior machinability, and enhanced thermal and electrical conductivities. On the other hand, they have high melting or decomposition temperatures and strong elastic stiffness, similar to many ceramics[5]. The unique combination of metallic and ceramic properties makes them versatile, with applications ranging from high-temperature coatings to nuclear accident-tolerant fuel (ATF), concentrated solar power (CSP), catalysis, and as precursors for MXenes [6], [7], [8], [9].

The diversity of MAX phases was confined to C and N as X elements for years (from 1960 to 2014). However, recent advancements have overcome this limitation by successfully synthesizing MAX phases that contain B [10]. The physical and chemical properties of B and B-containing compounds highlight the potential of MAX phase borides, which replace C/N with boron [11]. Due to the presence of B in the composition, they are also called the MAB phase. Furthermore, scientific communities are actively working to broaden the diversity of MAX phases by introducing structural changes, as seen in examples like $Cr_3AlB_4$ (space group *Immm*), $Cr_4AlB_6$ (space group Cmmm), $Cr_4AlB_4$ [12], [13], [14]. Khazaei *et al.* reported on the first investigation of the theoretical MAX phase borides $M_2AlB$ (where M = Sc, Ti, Cr, Zr, Nb, Mo, Hf, or Ta). They evaluated these compounds' electronic structure, mechanical characteristics, and dynamical stability in their investigation [15]. The diversity of MAX phases is a subject of research interest due to their structure variations and the number of atoms in the compounds, resulting in changes to their characteristics. In examples like the 211 MAX phase [16], 312 MAX phase [8][17], 212 MAX phase [3], and i-MAX phase [18], the recent additions to the MAX family include the 314 MAX phase and 212 MAX phase, where the element B has been incorporated as an X element [3] [19]. This introduces a novel aspect to the MAX phase family, contributing to its diversity. The unique structural features of MAB phases include both orthorhombic and hexagonal symmetry observed in crystals. These distinctive symmetries set MAB phases apart from the typical MAX phases. Yinqiao Liu's synthesis of the orthorhombic phase of $M_2AlB_2$, particularly the 212-MAB phases with M = Sc, Ti, Zr, Hf, V, Nb, Cr, Mo, W, Mn, Tc, Fe, Co, and Ni, has revealed unique structural stability and notable electrical and mechanical properties [20]. The MAB phase structure slightly differs from the standard MAX phases, which typically crystallize in the hexagonal system with a space group *P-6m$_2$* (No. 187). Ali *et al* [3]. investigated the diverse physical properties of $Zr_2AB_2$ (A = In, Tl). Martin Ade *et al* [13] synthesized ternary borides, namely $Cr_2AlB_2$, $Cr_3AlB_4$, and $Cr_4AlB_6$, and subsequently compared their mechanical properties. Qureshi *et al*. [19] investigated the 314 $Zr_3CdB_4$ MAX phase boride, calculating its mechanical, thermodynamic, and optical properties. The 314 MAX phase $Hf_3PB_4$ has been thoroughly studied using Density Functional Theory (DFT) and revealed that it was the hardest MAX phase compound discovered until that date [21].

The structure of 212 phases exhibits a slight deviation from 211 MAX phases. In the case of 212 phases, a 2D layer of B is situated between M layers, featuring an additional B atom at the X position, unlike in 211 phases [22] [23], [24]. The B-B bonding in 212 MAX phase borides has improved mechanical and thermal properties. To date, the physical properties of $Zr_2AB_2$ (A = In, Tl) [25] and $Hf_2AB_2$ (A = In, Sn) [26], $M_2AB$ (M= Ti, Zr, Hf; A=Al, Ga, In) [27] MAX phases, as well as $Nb_2AC$ (A = Ga, Ge, Tl, Zn, P, In, and Cd) MAX phases, have been investigated using density functional theory (DFT). In each instance, the mechanical properties of B-containing compounds show significant improvement compared to their traditional C/N containing 211 MAX phases.

The Debye temperature and melting temperature are higher for boron-containing 212 phases than for 211 carbides/nitrides, while the minimum thermal conductivity is lower. The thermal expansion coefficient of borides remains well-suited for use as coating materials. Consequently, the superior thermomechanical properties of B-containing 212 MAX phases demonstrate their suitability for high-temperature technological applications, surpassing the commonly used 211 MAX phase carbides. It should be noted that the 212 MAB phase with a hexagonal structure has already been synthesized [24]. In addition, Ga and Ge-based MAX phases have also been synthesized previously [28]. Thus, the reports on the synthesis of 212 phase and Ga and Ge-based MAX phases motivated us to select the 212 MAB phases: $M_2AB_2$ (M = Mo, Ta; A = Ga, Ge) for our present study, and we have performed an in-depth investigation of their physical properties through DFT method.

Therefore, in this paper, the first-time prediction of the stability and mechanical, electronic, thermal, and optical properties of $M_2AB_2$ (M = Mo, Ta; A = Ga, Ge) phases has been presented. The results revealed that $M_2AB_2$ (M = Mo, Ta; A = Ga, Ge) compounds are stable and suitable for thermal barrier coating (TBC) and reflection coating applications. Additionally, to provide a comparison, the properties determined for $M_2AB_2$ (M = Mo, Ta, and A = Ga, Ge) are compared with those of other 212 and 211 compounds MAX phase borides.

## 2. Methods of calculations

First principles density-functional theory (DFT) computations are conducted utilizing the Cambridge Serial Total Energy Package (CASTEP) module integrated within Materials Studio 2017 [29], [30]. The exchange-correlation function is estimated using the Generalized Gradient Approximation (GGA) method, originally suggested by Perdew, Burke, and Ernzerhof [31]. Pseudo-atomic simulations accounted for electronic orbitals corresponding to B ($2s^2\ 2p^1$), Ga ($3d^{10}\ 4s^2\ 4p^1$), Ge ($4s^2\ 4p^2$), Mo ($4s^2\ 4p^6\ 4d^5\ 5s^1$), and Ta ($5d^3\ 6s^2$). The energy cutoff and $k$-point grids were established at 650 eV and 11 × 11 × 4, respectively. The structural relaxation was performed utilizing the Broyden-Fletcher-Goldfarb-Shanno (BFGS) technique [32], while the electronic structure was computed employing density mixing. The parameters for relaxed structures incorporate the following tolerance thresholds: the self-consistent convergence of the total energy is set at $5 \times 10^{-6}$ eV/atom, the maximum force exerted on the atom is limited to 0.01 eV/Å, the maximum ionic displacement is constrained to $5 \times 10^{-4}$ Å, and a maximum stress threshold of 0.02 GPa is imposed. The finite strain method [33], grounded in density functional theory (DFT), is utilized to compute the elastic properties within this framework. All necessary equations for determining various properties are provided in the supplementary document.

## 3 Results and discussion

### 3.1 Structural properties

The $M_2AB_2$ compounds (where M=Mo or Ta; A=Ga or Ge) belong to the $P6m_2$ (No. 187)[23] space group and crystallize in the hexagonal system. Unlike conventional MAX phases, which typically belong to the $P6_3/mmc$ (194) space group, the 212 MAX phases exhibit distinct characteristics. In Fig. 1, the unit cell structure of $Mo_2GaB_2$ is depicted as a representative of $M_2AB_2$ alongside $Mo_2GaB$, facilitating a comparison to discern their differences easily. The atomic positions are as follows: M (Mo or Ta) at (0.3333, 0.6667, 0.6935), A at (0.6667, 0.3333, 0.0), and two B atoms positioned at (0.6667, 0.3333, 0.5) and (0.0, 0.0, 0.5). The B components are arranged at the corners of the unit cell in typical 211 MAX phases, but in 212 boride MAX phases, they form a 2D layer between the M layers. This structural arrangement results in B-B covalent bonds in the 2D layer, enhancing stability compared to conventional 211 MAX phases.

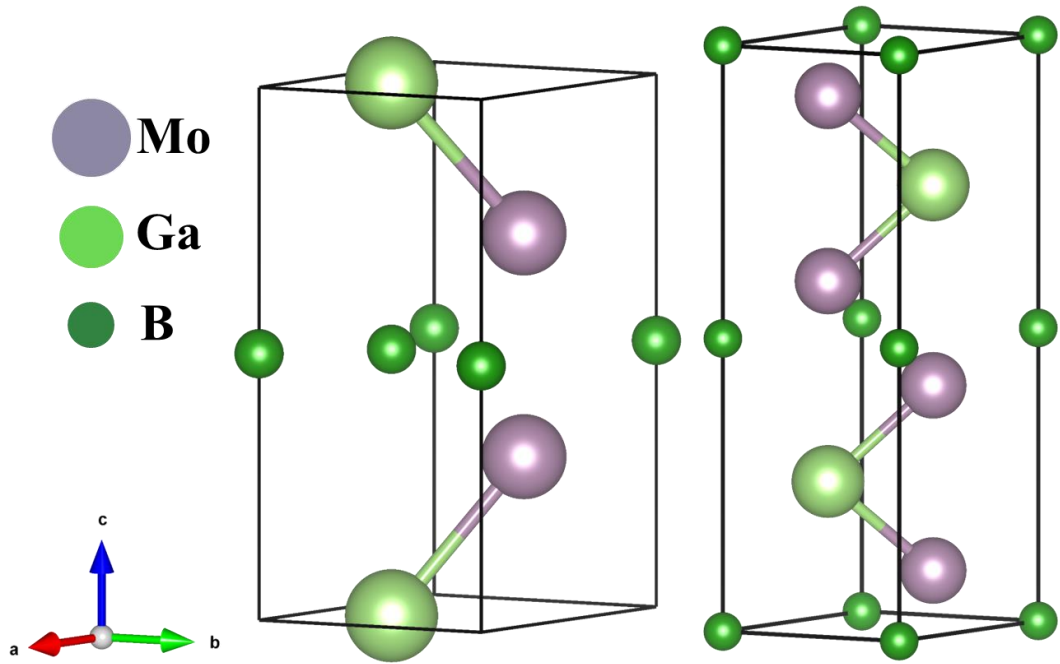

**Fig. 1** - The schematic unit cell of (a) $Mo_2GaB_2$ and (b) $Mo_2GaB$ compound.

The calculated lattice parameters of $M_2AB_2$ are presented in Table 1 alongside those of other 211 and 212 MAX phases for comparison, demonstrating consistency with prior results [16], [34], [35] and affirming the accuracy of the computational methodology used. The primary distinction between 212- and 211-unit cell structures arises from differences in the lattice parameter $c$, where the c value for 211 exceeds that of 212. Moreover, the volumes of Mo- and Ta-based 211 MAX phase borides surpass those of 212 MAX phase borides.

**Table 1-**Calculated lattice constants (*a* and *c*), *c/a* ratio, and volume (*V*) of $M_2AB_2$.

| Phase | $Mo_2GaB_2$ | $Mo_2GeB_2$ | $Ta_2GaB_2$ | $Ta_2GeB_2$ | [a]$Ti_2PB_2$ | [b]$Mo_2GaB$ | [c]$Ta_2GaB$ |
|---|---|---|---|---|---|---|---|
| *a* (Å) | 3.079 | 3.128 | 3.141 | 3.202 | 3.121 | 3.122 | 3.234 |
| *c* (Å) | 7.112 | 6.818 | 7.528 | 7.102 | 6.545 | 12.971 | 13.649 |
| *c/a* | 2.31 | 2.17 | 2.39 | 2.21 | 2.09 | 4.15 | 4.22 |
| *V* (Å³) | 58.41 | 57.77 | 64.34 | 63.07 | - | 109 | 123.6 |

[a]Reference [34], [b]Reference [16], [c]Reference [35].

### 3.2 Stability

Examining a compound's stability is significant for multiple purposes, as it yields valuable information regarding the compound's synthesis parameters and aids in assessing the material's resilience across diverse environments, including thermal, compressive, and mechanical pressures. In this section, we delve into a comprehensive theoretical analysis concerning the chemical, dynamic, and mechanical stability of $M_2AB_2$ compounds.

The compound's chemical stability is determined by computing its formation energy by the following equation [36]: $E_{for}^{M_2AB_2} = \frac{E_{total}^{M_2AB_2} - (xE_{solid}^M + yE_{solid}^A + zE_{solid}^B)}{x+y+z}$. In the context provided, $E_{total}^{M_2AB_2}$ represents the total energy of the compound after optimizing the unit cell. $E_{solid}^M$, $E_{solid}^A$, and $E_{solid}^B$ denote the energies of the individual elements M, A, and B, respectively. The variables *x*, *y*, and *z* correspond to the number of atoms in the unit cell for M, Ga, and B, respectively. For $Mo_2GaB_2$, $Mo_2GeB_2$, $Ta_2GaB_2$, and $Ta_2GeB_2$, the calculated formation energy ($E_f$) are -1.8079 eV/atom, -1.8859 eV/atom, -2.1252 eV/atom, and -2.1933 eV/atom, respectively. The negative values signify the chemical stability of all compounds. The order of chemical stability can be expressed as $Ta_2GeB_2 > Ta_2GaB_2 > Mo_2GeB_2 > Mo_2GaB_2$, indicating that $Ta_2GeB_2$ is the most stable. Additionally, it's observed that MAX phase borides containing Ta are more stable than those containing Mo.

Negative formation energy alone may not fully explain the chemical stability of $M_2AB_2$ (M=Mo, Ta and A=Ga, Ge). We calculated its formation enthalpy by examining potential pathways to evaluate its thermodynamic stability. For this analysis, we used the experimentally identified stable phases of MoB[37] and TaB[38], Ga$_4$Mo[39], and B$_2$Mo[40]. The potential decomposition pathways for our compounds, as determined from the Open Quantum Materials Database (OQMD), are outlined below.

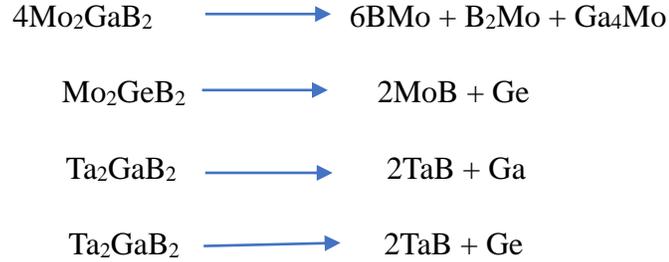

$$4Mo_2GaB_2 \longrightarrow 6BMo + B_2Mo + Ga_4Mo$$

$$Mo_2GeB_2 \longrightarrow 2MoB + Ge$$

$$Ta_2GaB_2 \longrightarrow 2TaB + Ga$$

$$Ta_2GaB_2 \longrightarrow 2TaB + Ge$$

We have calculated the reaction energy as follows [41]:

$$E_{reac} = \left(\Delta H_f^{nM2AB2}\right) - \left(\sum_{Stable\ phases} \Delta H_f^{products}\right)$$

Where, M=Mo, Ta and A=Ga, Ge.

The following formula can calculate the decomposition energy associated with the reaction energy. $E_{decom} = \frac{E_{reac}}{n}$. Where n is the number of participating atoms. The calculated decomposition energies for Mo$_2$GaB$_2$, Mo$_2$GeB$_2$, Ta$_2$GaB$_2$, and Ta$_2$GeB$_2$ are -0.16, -0.17, -0.26, and -0.28 meV/atom, respectively. So, we can state that the M$_2$AB$_2$ (M=Mo, Ta and A=Ga, Ge) system exhibits thermodynamic stability.

Phonon dispersion curves (PDCs) have been calculated at the ground state utilizing the density functional perturbation theory (DFPT) linear-response approach to evaluate the dynamic stability of the MAX phase borides under investigation [42]. The PDCs, depicting the phonon dispersion along the high symmetry directions of the crystal Brillouin zone (BZ), along with the total phonon density of states (PHDOS) of M$_2$AB$_2$ (M=Mo, Ta; A=Ga, Ge) compounds, are illustrated in Fig. 2(a, b, c and d). Analysis of the PDCs reveals no negative phonon frequencies for any of the compounds, indicating their dynamic stability. The PHDOS of the M$_2$AB$_2$ compounds are derived from the PDCs and are presented alongside the PDCs in Fig. 2(a, b, c, and d), facilitating band identification through comparison of corresponding peaks. From Fig. 2, it is observed that in Mo$_2$GaB$_2$, the flatness of the bands for the Transverse Optical (TO) modes results in a prominent peak in the PHDOS, whereas non-flat bands for the Longitudinal Optical (LO) modes lead to weaker peaks in the PHDOS. Similar trends are observed in Mo$_2$GeB$_2$, Ta$_2$GaB$_2$, and

Ta$_2$GeB$_2$. Notably, a distinct discrepancy arises between the optical and acoustic branches, with the top of the LO and bottom of the TO modes situated at the *G* point, with separations of 7.49 THz, 6.39 THz, 9.39 THz, and 8.51 THz for Mo$_2$GaB$_2$, Mo$_2$GeB2, Ta$_2$GaB$_2$, and Ta$_2$GeB$_2$ compounds, respectively.

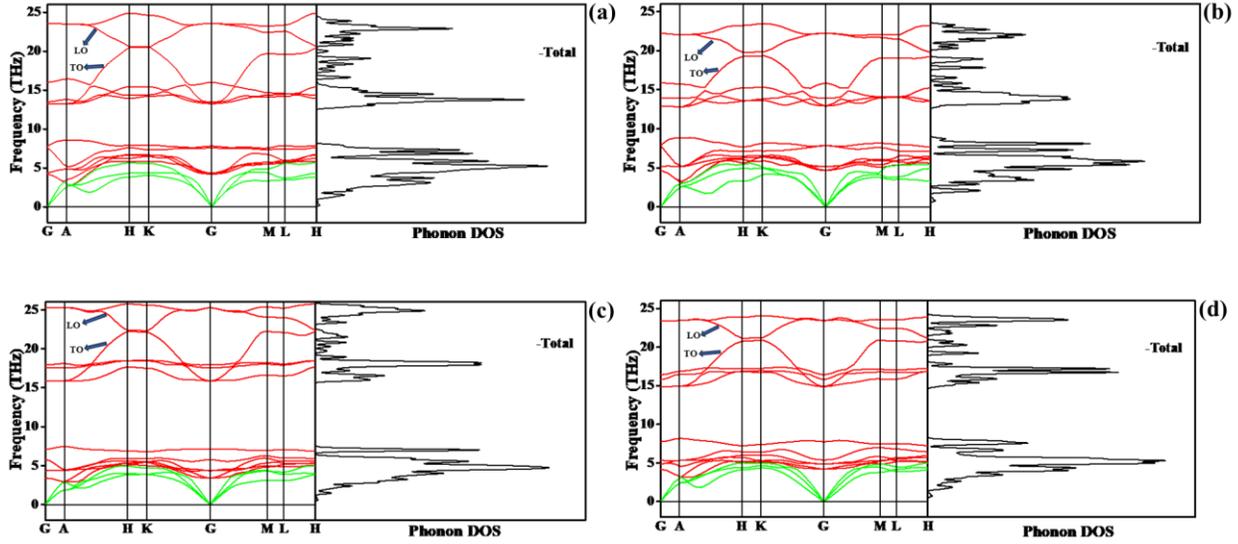

**Fig. 2** – Phonon DOS and Phonon dispersion curves of (a) Mo$_2$GaB$_2$, (b) Mo$_2$GeB$_2$, (c)Ta$_2$GaB$_2$, and (d)Ta$_2$GeB$_2$ compounds.

Materials are subjected to various forces and loads in practical applications, necessitating understanding their mechanical stability. The mechanical stability of a compound can be assessed using stiffness constants. For a hexagonal system, the conditions for mechanical stability are as follows [43]: $C_{11} > 0$, $C_{11} > C_{12}$, $C_{44} > 0$, and $(C_{11} + C_{12})C_{33} - 2(C_{13})^2 > 0$. As indicated in Table 2, the $C_{ij}$ values of M$_2$AB$_2$ (M=Mo, Ta; A=Ga, Ge) satisfy these conditions, thus confirming the mechanical stability of herein predicted phases: M$_2$AB$_2$ (M=Mo, Ta; A=Ga, Ge).

### 3.3 Electronic properties

Analyzing the electronic band structure (EBS) is crucial for gaining insights into the electronic behavior of a compound. The EBS of M$_2$AB$_2$ (M=Mo, Ta; A=Ga, Ge) MAX phases are depicted in Fig. 3(a, b, c, and d), with the Fermi energy (E$_F$) level set at 0 eV, represented by a horizontal line. Observing the EBS of M$_2$AB$_2$ (M=Mo, Ta; A=Ga, Ge), it is evident that the conduction band overlaps with the valence band, indicating the absence of a band gap. This observation confirms that the M$_2$AB$_2$ compounds exhibit metallic behavior, which aligns with conventional

MAX phases. The red lines illustrate the overlapping band at the Fermi level. Fig. 3 (a, b) shows that for Mo$_2$GaB$_2$ and Mo$_2$GeB$_2$ compounds, the maximum band overlap occurs along the *A-H* path. Conversely, in Fig. 3(c, d), the maximum band overlap is observed along the *G-M* paths. Utilizing the band structure, we can analyze the electrical anisotropy of M$_2$AB$_2$ MAX phase compounds. The anisotropic nature can be understood by analyzing the energy dispersion in the basal plane and along the c-axis. The paths *G-A*, *H-K*, and *M-L* show energy dispersion along the *c*-direction, and *A-H*, *K-G*, *G-M*, and *L-H* show energy dispersion in the basal plane. In comparison to the paths *A-H*, *K-G*, *G-M*, and *L-H* (basal plane), there is less energy dispersion along the lines *G-A*, *H–K*, and *M-L* (c-direction), as shown by Fig. 2(a, b, c and d). Lower energy dispersion in the *c*-direction results from a higher effective mass [44], indicating the strong electronic anisotropy of the M$_2$AB$_2$ MAX phase compound. Consequently, conductivity along the *c*-axis is expected to be lower than in the basal planes. These findings are consistent with prior studies [34], [45].

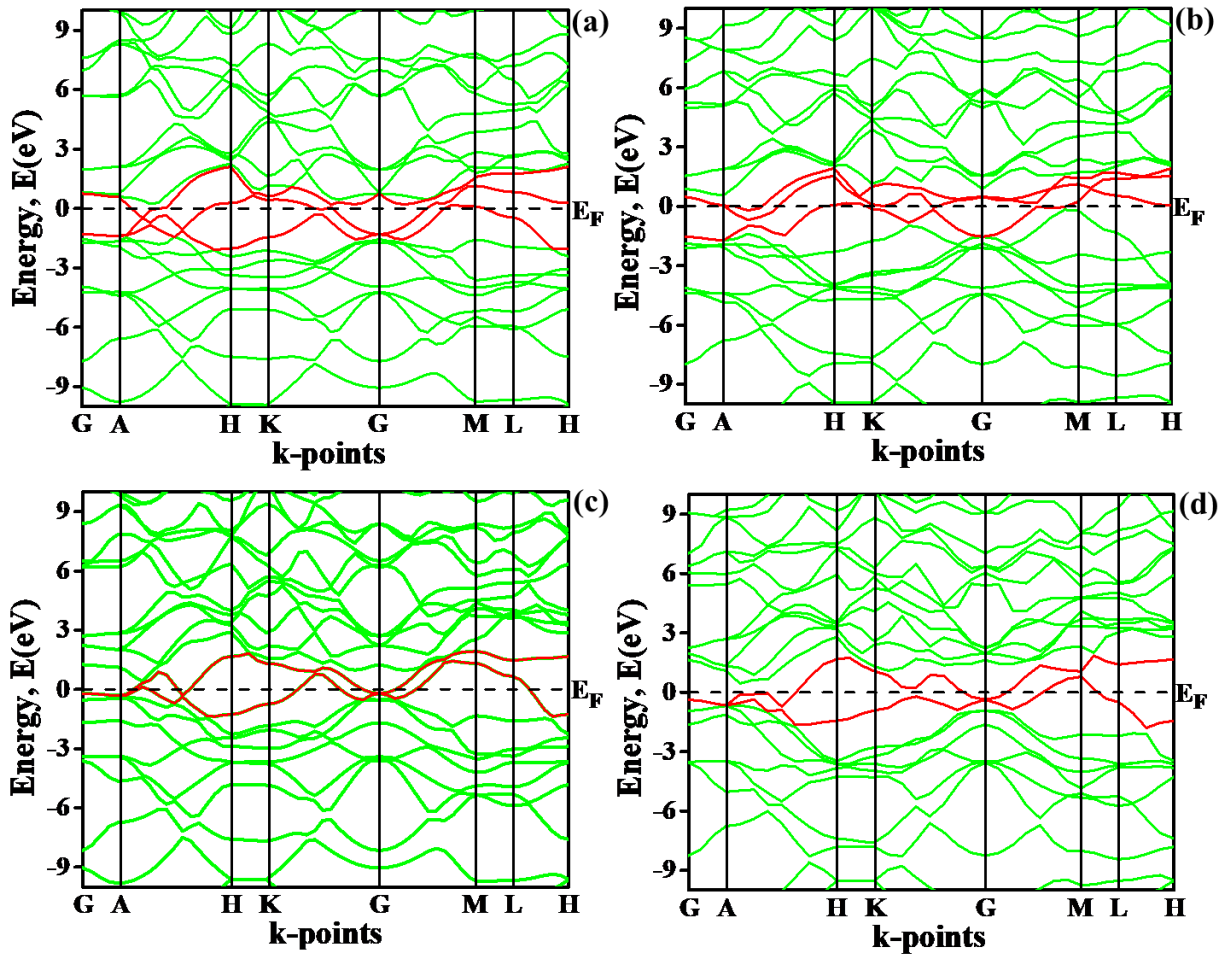

**Fig. 3** – Band structure of (a) $Mo_2GaB_2$, (b) $Mo_2GeB_2$, (c) $Ta_2GaB_2$, and (d) $Ta_2GeB_2$ compounds.

To investigate the bonding nature and electronic conductivity, the total density of states (TDOS) and partial density of states (PDOS) of $M_2AB_2$ compounds were computed. Figure 4 (a, b, c, and d) illustrates the TDOS and PDOS of these compounds, with the Fermi energy ($E_F$) set at zero energy level, indicated by a straight line. These profiles exhibit typical characteristics of MAX phase materials. The Mo or Ta-*d* electronic states predominantly contribute to the Fermi level, with a minor contribution from the B-*p* and Ga or Ge-*p* electronic states. To ascertain the hybridization characteristics of various electronic states within the valence band, the energy spectrum of the valence band has been partitioned into two distinct segments. The first segment encompasses the lower valence band region spanning from -7 eV to -3.5 eV, originating from the hybridization of Mo-*p*, Mo-*d*, and B-*s* orbitals in the case of the $Mo_2GaB_2$ compound. Conversely, for the $Mo_2GeB_2$ compound, the lower valence band region arises from the hybridization of Mo-*d*, Ge-*p*, and B-*s* orbitals. Notably, for both the $Ta_2GaB_2$ and $Ta_2GeB_2$ compounds, the dominance of the B-*p* state characterizes the lower valence band region. The second segment pertains to the upper valence band region from -3.5 eV to 0 eV. In the case of the $Mo_2GaB_2$ compound, this region arises from the hybridization of Mo-p and Mo-d orbitals. However, for the $Mo_2GeB_2$, $Ta_2GaB_2$, and $Ta_2GeB_2$ compounds, the upper valence band region results from the hybridization of Mo-*d* orbitals and (Ga/Ge)-*p* orbitals. Notably, the Fermi level of $M_2AB_2$ resides near the pseudogap in the TDOS profile, indicating a high level of electronic stability. Similar trends are observed in other compounds like $Zr_2AlN$, $V_2AlN$, $Sc_2AlB$, $Sc_2GaB$, $Ta_2GaB$, and $Hf_2GaB_2$ [15], [35], [45].

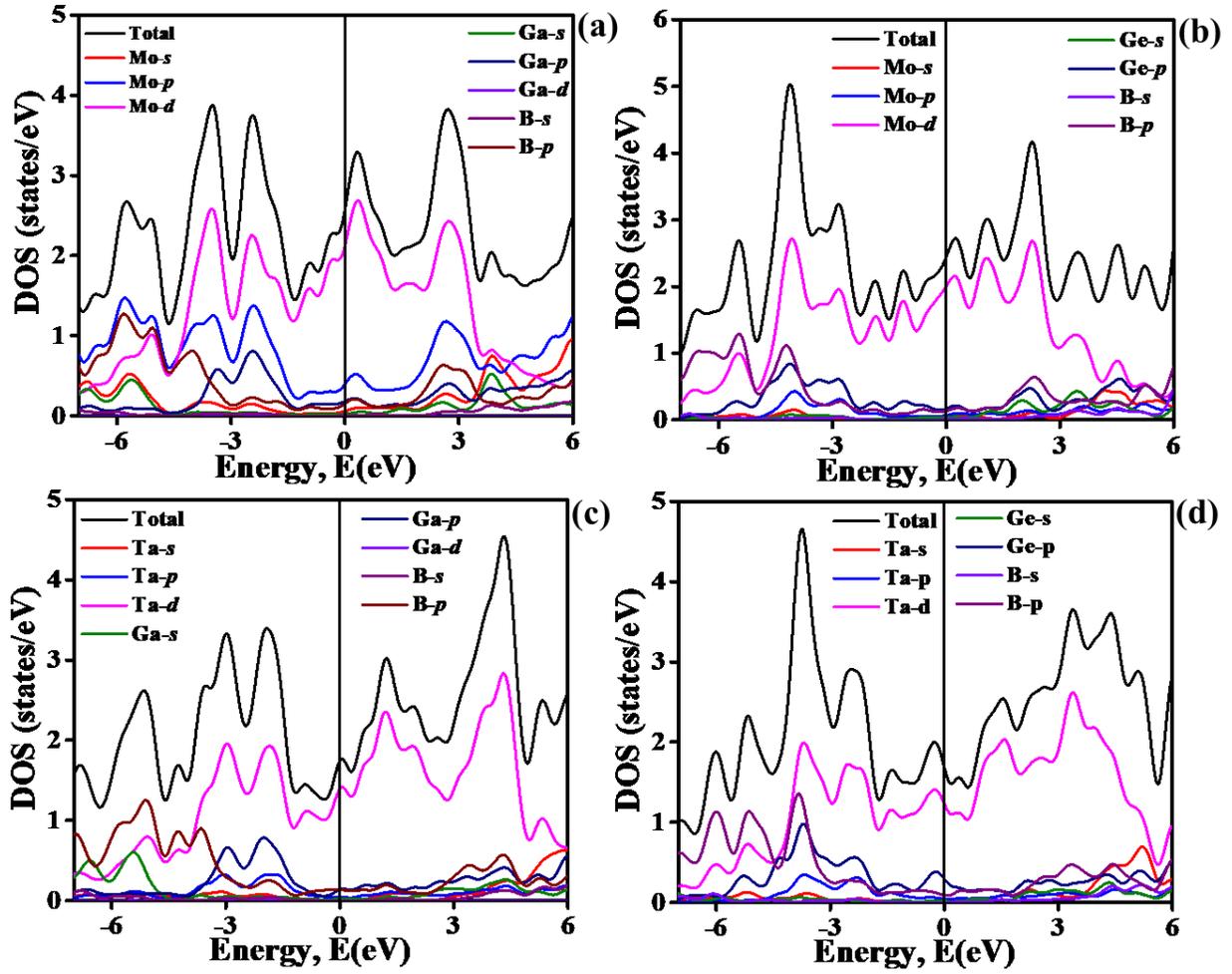

**Fig. 4** – TDOS and PDOS of (a) $Mo_2GaB_2$, (b) $Mo_2GeB_2$, (c) $Ta_2GaB_2$, and (d) $Ta_2GeB_2$ compounds.

The electron charge density mapping is helpful in understanding the distribution of electron densities linked to chemical bonds. It delineates areas of positive and negative charge densities, signifying the development and exhaustion of electrical charges, respectively. As depicted in the map, covalent bonds become apparent by accumulating charges between two atoms. Furthermore, the presence of ionic bonds can be inferred from a balance between negative and positive charges at specific atom positions[46]. The valence electronic CDM, denoted in units of eÅ$^{-3}$, for $M_2AB_2$ (where M = Mo, Ta; A = Ga, Ge) is showcased in Fig. 5(a, b, c, and d) along the (110) crystallographic plane. The accompanying scale illustrates the intensity of electronic charge density, with red and blue colors indicating low and high electronic charge density, respectively. As depicted in Fig. 5(a, b, c, and d), it is evident that charges accumulate in the regions between the B sites. Consequently, it is anticipated that strong covalent B–B bonding occurs through the formation of two center-two electron (2c–2e) bonds in the $M_2AB_2$ (where M=Mo, Ta; A=Ga, Ge) compound, similar to other 212 MAX phase borides like $Ti_2PB_2$, $Zr_2PbB_2$, $Nb_2SB_2$, $Zr_2GaB_2$ and $Hf_2GaB_2$ [34], [45]. Mulliken analysis has corroborated the charge transfer from Mo/Ta atoms to B atoms. The charge received from Mo/Ta atoms is distributed among the B atoms positioned at the transitions and those located at the edges, facilitating the formation of a two center-two electron (2c–2e) bond between B atoms within the 2D layer of B, as illustrated in Fig. 1(a). The hardness of each bond value presented in Table 5 also aligns with the results obtained from charge density mapping (CDM) and our analysis using moduli and elastic stiffness constants.

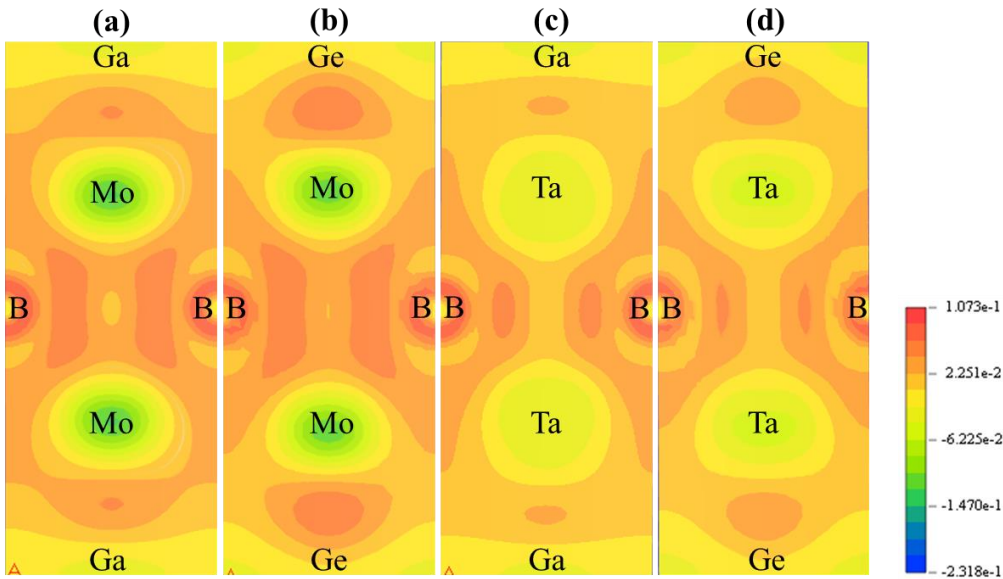

**Fig. 5** – The charge density mapping of (a) $Mo_2GaB_2$, (b) $Mo_2GeB_2$, (c) $Ta_2GaB_2$, and (d) $Ta_2GeB_2$ compounds.

### 3.4.1 Mechanical properties

The mechanical properties of materials play a pivotal role in determining their potential applications, serving as crucial indicators of their behavior and suitability in materials engineering endeavors. These properties are equally applicable to MAX phase materials. Initially, to assess the mechanical properties of the $M_2AB_2$ (M=Mo, Ta; A=Ga, Ge) phases, we employed the stress-strain method within the CASTEP code to compute the elastic constants ($C_{ij}$) [22], [47], [48]. These calculated elastic constants ($C_{ij}$) are presented in Table 2, alongside those of other 212 and 211 MAX phases. Due to the hexagonal crystal structure of M$_2$AB$_2$ (M=Mo, Ta; A=Ga, Ge) phase borides, five stiffness constants emerge: $C_{11}$, $C_{12}$, $C_{13}$, $C_{33}$, and $C_{44}$ [49]. The mechanical stability was evaluated using these stiffness constants in the preceding section. For instance, $C_{11}$ and $C_{33}$ determine stiffness when stress is exerted along the (100) and (001) directions, respectively, whereas $C_{44}$ evaluates resistance to shear deformation on the (100) and (001) planes. The stiffness constants $C_{11}$ and $C_{12}$ directly reflect the strength of atomic bonds along the *a*- and *c*-axes. When $C_{11}$ exceeds $C_{33}$ (or vice versa), it signifies stronger atomic bonding along the *a*-axis (or *c*-axis). In Table 2, for Mo$_2$GaB$_2$, Mo$_2$GeB$_2$, and Ta$_2$GeB$_2$ compounds, $C_{11}$ surpasses $C_{33}$, indicating superior atomic bonding along the *a*-axis compared to the *c*-axis. This robust bonding along the *a*-axis suggests heightened resistance to *a*-axial deformation. Conversely, in Ta$_2$GeB$_2$, where $C_{11} < C_{33}$, stronger atomic bonding along the *c*-axis translates to increased resistance against c-axial deformation. Analysis of Table-2 reveals that the values of $C_{11}$ and $C_{22}$ are notably higher for 212 MAX phase borides than 211 MAX phase borides. Consequently, it can be inferred that 212 MAX phases exhibit stronger resistance to axial deformation when juxtaposed with 211 MAX phases. $C_{44}$ is commonly utilized to gauge shear deformation tolerance among the elastic constants. Notably, Ta$_2$GaB exhibits superior shear deformation resistance owing to its highest $C_{44}$ value among the studied compounds. Furthermore, the individual elastic constants $C_{12}$, $C_{13}$, and $C_{44}$ denote shear deformation response under external stress. The observation that $C_{11}$ and $C_{33}$ possess larger magnitudes than $C_{44}$ implies that shear deformation is more facile than axial strain. Another crucial parameter derived from the stiffness constants is the Cauchy pressure (*CP*), calculated as $C_{12} - C_{44}$, which provides vital insights relevant to the practical applications of solids [51]. Pettifor [51] emphasized the significance of Cauchy pressure (*CP*) in discerning the chemical bonding and ductile/brittle properties of solids. A negative *CP* value indicates covalently bonded brittle solids, whereas a positive value signifies isotropic ionic ductile solids. $Mo_2GaB_2$ and $Ta_2GaB_2$ fall into covalently bonded brittle solids, while $Mo_2GeB_2$ and $Ta_2GeB_2$ exhibit metallic ductile characteristics. Similarly, like Ga-containing 211 MAX phase borides, Ga-containing 212 MAX phases also demonstrate negative *CP* values and brittle behavior [52][53].

**Table 2-** The stiffness constants, elastic moduli, Cauchy Pressure (*CP*), *f*-index, Pugh's ratio (*G/B*), fracture toughness ($K_{IC}$), and hardness parameters of M$_2$AB$_2$.

| Parameters | Mo$_2$GaB$_2$ | Mo$_2$GeB$_2$ | Ta$_2$GaB$_2$ | Ta$_2$GeB$_2$ | [a]Ti$_2$PB$_2$ | [b]Ti$_2$InB$_2$ | [c]Zr$_2$TlB$_2$ | [d]Mo$_2$GaB | [e]Ta$_2$GaB |
|---|---|---|---|---|---|---|---|---|---|
| $C_{11}$ (GPa) | 393 | 385 | 339 | 334 | 368 | 358 | 310 | 312 | 309 |
| $C_{33}$ (GPa) | 322 | 372 | 332 | 395 | 383 | 270 | 251 | 246 | 278 |
| $C_{44}$ (GPa) | 149 | 161 | 145 | 178 | 184 | 93 | 66 | 158 | 141 |
| $C_{12}$ (GPa) | 125 | 164 | 128 | 186 | 90 | 52 | 52 | 95 | 107 |
| $C_{13}$ (GPa) | 146 | 167 | 133 | 146 | 121 | 57 | 61 | 167 | 115 |
| $B_V$ (GPa) | 216 | 238 | 199 | 224 | - | - | - | 192 | 174 |
| $B_R$ (GPa) | 215 | 238 | 199 | 224 | - | - | - | 192 | 174 |
| $B$ (GPa) | 215 | 238 | 199 | 224 | 198 | 145 | 135 | 192 | 174 |

| $G_V$ (GPa) | 132 | 129 | 120 | 125 | - | - | - | 114 | 114 |
| $G_R$ (GPa) | 129 | 124 | 117 | 107 | - | - | - | 87 | 109 |
| $G$ (GPa) | 130 | 127 | 118 | 116 | 152 | 119 | 94 | 101 | 111 |
| $Y_V$ (GPa) | 330 | 329 | 300 | 316 | - | - | - | 286 | 281 |
| $Y_R$ (GPa) | 322 | 318 | 293 | 278 | - | - | - | 228 | 271 |
| $Y$ (GPa) | 326 | 323 | 297 | 297 | 361 | 280 | 229 | 258 | 276 |
| $v_V$ | 0.245 | 0.269 | 0.249 | 0.265 | - | - | - | 0.252 | 0.232 |
| $v_R$ | 0.249 | 0.276 | 0.255 | 0.293 | - | - | - | 0.302 | 0.241 |
| $v$ | 0.247 | 0.273 | 0.252 | 0.279 | 0.19 | - | 0.22 | 0.276 | 0.236 |
| $CP$ | -36.10 | 2.66 | -17.24 | 8.16 | -94 | - | -21 | - | -33.96 |
| $G/B$ | 0.60 | 0.53 | 0.59 | 0.51 | 0.77 | 0.821 | .70 | 0.53 | 0.64 |
| $KIC$ (MPa.m$^{1/2}$) | 2.53 | 2.62 | 2.36 | 2.47 | - | - | - | 2.16 | 2.21 |
| $f$ | 1.284 | 1.045 | 1.010 | 0.918 | - | - | - | 0.937 | 1.142 |
| $H_{macro}$ (GPa) | 16.32 | 13.35 | 14.76 | 11.98 | 24.74 | - | 15.7 | 11.051 | 15.75 |
| $H_{micro}$ (GPa) | 22.04 | 19.23 | 19.61 | 17.14 | 30.79 | - | 17.7 | 15.098 | 19.68 |

[a]Reference [34], [b]Reference [50], [c]Reference [25], [d]Reference [16], [e]Reference [35].

The elastic constants obtained are used to produce several bulk elastic parameters that are used to characterize polycrystalline materials, such as Young's modulus (*Y*), bulk modulus (*B*), and shear modulus (*G*). In Table 2, the bulk modulus (*B*) and shear modulus (*G*) calculated using Hill's approximation [54] are also presented. Hill's values represent the average of the upper limit (Voigt [55]) and lower limit (Reuss [56]) of *B*. The necessary equations for these calculations are provided in the supplementary document (S1). Young's modulus (*Y*) is a crucial indicator of material stiffness. A higher *Y* value indicates a stiffer material. It can be observed from the table that $Mo_2GaB_2$ exhibits the highest *Y* value, signifying its greater stiffness compared to others.

According to the sequence of $Y$ values, stiffness can be ranked as follows: $Mo_2GaB_2$ > $Mo_2GeB_2$ > $Ta_2GaB_2$ > $Ta_2GeB_2$. Therefore, $Mo_2GaB_2$, with its higher Young's modulus, is anticipated to demonstrate superior mechanical stability and deformation resistance compared to the other compounds under investigation. This quality is essential for aircraft parts or high-performance machinery applications where critical dimensional stability and structural integrity are required [57], [58]. Table 2 shows that 212 MAX phase materials exhibit larger Young's modulus ($Y$) values than 211 MAX phases. This suggests that 212 MAX phase borides are stiffer than their 211 MAX phase counterparts. Young's modulus ($E$) also correlates well with thermal shock resistance ($R$): $R \propto 1/E$ [59]. Lower Young's modulus values correspond to higher thermal shock resistance. Therefore, materials with higher thermal shock resistance (i.e., lower Young's modulus) are more suitable for use as Thermal Barrier Coating (TBC) materials. Given that $Ta_2GaB_2$ possesses the lowest $Y$ value among the materials studied, it should be considered a superior candidate for TBC material due to its higher thermal shock resistance. The material's ability to withstand shape distortion is elucidated by the shear modulus ($G$). On the other hand, the bulk modulus ($B$) indicates the strength of a material's chemical bonds and its ability to withstand uniform compression or volume change. We computed Young's modulus ($Y$), bulk modulus ($B$), and shear modulus for our analysis. Based on the bulk modulus values presented in Table 2, the sequence of a material that resists compression when pressure is applied can be outlined as follows: $Mo_2GeB_2$ > $Ta_2GeB_2$ > $Mo_2GaB_2$ > $Ta_2GeB_2$. $Mo_2GeB_2$, boasting a higher bulk modulus, may exhibit reduced plastic deformation and superior resistance to stress-induced deformation compared to other compounds examined.

The ductile and brittle characteristics of the $M_2AB_2$ (M=Mo, Ta; A=Ga, Ge) compounds can also be assessed through Poisson's ratio ($v$) and Pugh's ratio ($G/B$). A compound demonstrates ductile (brittle) behavior if the $v$ value surpasses (falls below) 0.26 [60]. Furthermore, if the $G/B$ value exceeds (is less than) 0.571, then the compound exhibits brittle (ductile) behavior [19]. According to both criteria, $Mo_2GaB_2$ and $Ta_2GaB_2$ are classified as brittle compounds, whereas $Mo_2GeB_2$ and $Ta_2GeB_2$ are categorized as ductile compounds. The ductile nature of Ge-based MAX phases has been reported previously [61].

Fracture toughness ($K_{IC}$) is a vital property that gauges a material's ability to resist crack propagation. In the case of $M_2AB_2$ (M=Mo, Ta; A=Ga, Ge) compounds, Equation (S2) was employed to determine $K_{IC}$. The $K_{IC}$ values are documented in Table 2, and notably, these values surpass those of 211 MAX phases [16], [35]. Another parameter, the "*f*-index," characterizes the isotropic nature and strength of atom-atom bonds within a single hexagonal crystalline lattice along the a- and c-directions. If the *f*-index is less than 1, chemical bonds exhibit greater rigidity along the *c*-axis; conversely, if *f* exceeds 1, bonds are more rigid in the *ab*-plane. When the value of *f* is set to 1, atomic bonds exhibit similar strength and uniformity in all directions [15]. The *f*-value is computed using Equation (S3) and displayed in Table 2. Table 2 shows that the *f*-values of $Mo_2GeB_2$, $Ta_2GaB_2$, and $Ta_2GeB_2$, which are close to one, indicate a slight anisotropic bonding strength. Nevertheless, substances with robust bonds in the horizontal plane (*ab* plane) (*f* > 1), such as $Mo_2GaB_2$, are deemed optimal candidates for the exfoliation process. In engineering applications, the hardness of a solid material serves as a valuable criterion for designing various devices. The elastic properties of polycrystalline materials can be utilized to calculate hardness values, as the ability to resist indentation is closely linked to a material's hardness. Both micro-hardness ($H_{micro}$) and macro-hardness ($H_{macro}$) were computed using Equation (S4) and are presented in Table 2. Based on the values of $H_{micro}$ and $H_{macro}$, $Mo_2GaB_2$ emerges as the toughest among the studied phases. The order of hardness is as follows: $Mo_2GaB_2$ > $Ta_2GaB_2$ > $Mo_2GeB_2$ > $Ta_2GeB_2$. Interestingly, Ga-containing MAX phase borides exhibit superior hardness compared to Ge-containing MAX phase borides. In Table 2, we compare our findings with previously reported MAX phases and observe that our data align perfectly with the earlier results.

### 3.4.2 Elastic anisotropy

Additionally, anisotropy is linked to other crucial events like anisotropic plastic deformation and the development and spread of micro-cracks within mechanical stress. By providing direction-dependent elastic constants, the understanding of anisotropy also offers a framework for improving the mechanical stability of materials in extreme circumstances. The following formulae are used for hexagonal structures to calculate the various anisotropic variables from elastic constants $C_{ij}$ [62].

$$A_1 = \frac{\frac{1}{6}(C_{11}+C_{12}+2C_{33}-4C_{13})}{C_{44}}; \; A_2 = \frac{2C_{44}}{C_{11}-C_{12}} \text{ and } A_3 = A_1.A_2 = \frac{\frac{1}{3}(C_{11}+C_{12}+2C_{33}-4C_{13})}{C_{11}-C_{12}}$$

Table 3 lists every anisotropy parameter. Since the value of $A_i$ should be 1 to be isotropic, the computed value of $A_i$ ($i = 1\text{-}3$) indicates that all of the compounds under research exhibit anisotropic behavior [63].

**Table 3:** Data for Elastic anisotropy factors.

| Phase | $A_1$ | $A_2$ | $A_3$ | $A_B$ | $A_G$ | $A^U$ |
|---|---|---|---|---|---|---|
| $Mo_2GaB_2$ | 0.65 | 1.12 | 0.73 | 0.22 | 1.35 | 0.14 |
| $Mo_2GeB_2$ | 0.64 | 1.46 | 0.93 | 0.006 | 1.92 | 0.196 |
| $Ta_2GaB_2$ | 0.69 | 1.38 | 0.95 | 0.0003 | 1.41 | 0.143 |
| $Ta_2GeB_2$ | 0.68 | 2.4 | 1.63 | 0.02 | 7.56 | 0.818 |
| [a]$Hf_2InB_2$ | 1.16 | 0.67 | 0.78 | - | - | 0.177 |
| [b]$Hf_2SnB_2$ | 1.04 | 0.76 | 0.79 | - | - | 0.040 |
| [c]$Zr_2PbB_2$ | 1.22 | 0.63 | 0.77 | - | - | 0.45 |

[a,b] Reference [64], [c] Reference [65].

An additional method for estimating elastic anisotropy is to use the percentage anisotropy to compressibility and shear ($A_B$ & $A_G$). This gives polycrystalline materials a helpful way to measure elastic anisotropy. They have been described as [45];

$$A_B = \frac{B_V - B_R}{B_V + B_R} \times 100\% \;;\; A_G = \frac{G_V - G_R}{G_V + G_R} \times 100\% \text{ and } A^U = 5\frac{G_V}{G_R} + \frac{B_V}{B_R} - 6 \geq 0$$

Zero values for AB, AG, and the universal anisotropy factor (AU) show elastic isotropy; the maximum amount of anisotropy is represented by a value of 1. Table 3 shows that, compared to other compounds, the values of $A_B$, $A_G$, and $A^U$ for $Mo_2GeB_2$ and $Ta_2GaB_2$ are incredibly close to zero, suggesting that these compounds have nearly isotropic characteristics.

### 3.4.3 Mulliken Populations

The Mulliken charge assigned to an atomic species quantifies the effective valence by calculating the absolute difference between the formal ionic charge and the Mulliken charge. Equations (S5) and (S6) are employed to ascertain the Mulliken charge for each atom (α). Table-4 provides the Mulliken atomic population and effective valence charge. Transition metals Mo and Ta in $M_2AB_2$ (M = Mo, Ta, and A = Ga, Ge) have pure valence states of $4d^5$ and $5d^3$, respectively. The *d*- orbital electrons of transition metals have been found to influence their effective valence charge significantly. A non-zero positive value indicates a combination of covalent and ionic attributes within chemical bonds. As this value decreases towards zero, it signifies a rise in ionicity. A zero value suggests an ideal ionic character in the bond. Conversely, a progression from zero with a positive value indicates an elevation in the covalency level of the bonds. Based on their effective valence, M atoms move from the left to the right in the periodic table, increasing the covalency of $M_2AB_2$ (M = Mo, Ta, and A = Ga, Ge). Table-4 reveals that the Mulliken atomic charge ascribed to the B atoms is solely negative. Conversely, positive Mulliken atomic charges are associated with transition metals (M) and A. This suggests a charge transfer from M and A to B for each compound within $M_2AB_2$ (where M = Mo, Ta, and A = Ga, Ge), thereby fostering ionic chemical bonds among these atoms. Bond population serves as another indicator of bond covalency within a crystal, as a high value of bond population essentially signifies a heightened degree of covalency within the chemical bond. The bonding and anti-bonding states influence the populations with positive and negative bond overlap. As demonstrated in Table-5, the B-B bond exhibits greater covalency compared to any other bond in $M_2AB_2$ (M = Mo, Ta, and A = Ga, Ge). The presence of an antibonding state between two relevant atoms, which subsequently decreases their chemical bonding, is indicated by a hostile bond overlap population. In $Mo_2GeB_2$ and $Ta_2GeB_2$, a hostile bond overlap population is observed in the Ge-Mo and Ge-Ta bonds, indicating the presence of an antibonding state. Therefore, the existence of ionic bonding is guaranteed by electronic charge transfer. On the other hand, the high positive value of the bond overlap population (BOP) denotes the presence of covalent bonding, a feature shared by materials in the MAX phase.

**Table 4-**Data for Mulliken atomic populations and Effective valence charge of $M_2AB_2$ (M = Mo, Ta, and A = Ga, Ge) compounds.

| Compound | Atom | Mulliken atomic population | | | | | Effective Vallance charge |
|---|---|---|---|---|---|---|---|
| | | s | p | d | Total | Charge | |
| | B | 0.89 | 2.57 | 0.00 | 3.46 | -0.46 | _ |
| | B | 0.90 | 2.58 | 0.00 | 3.48 | -0.48 | _ |
| **Mo₂GaB₂** | Ga | 0.69 | 1.84 | 9.99 | 12.52 | 0.48 | 2.52 |
|  | Mo | 2.21 | 6.48 | 5.08 | 13.77 | 0.23 | 5.77 |
|  | Mo | 2.21 | 6.48 | 5.08 | 13.77 | 0.23 | 5.77 |
| Mo$_2$GeB$_2$ | B | 0.90 | 2.55 | 0.00 | 3.45 | -0.45 | _ |
|  | B | 0.91 | 2.56 | 0.00 | 3.45 | -0.47 | _ |
|  | Ge | 0.77 | 2.50 | 0.00 | 3.27 | 0.73 | 3.27 |
|  | Mo | 2.28 | 6.50 | 5.12 | 13.91 | 0.09 | 5.91 |
|  | Mo | 2.28 | 6.50 | 5.12 | 13.91 | 0.09 | 5.91 |
| Ta$_2$GaB$_2$ | B | 1.01 | 2.58 | 0.00 | 3.59 | -0.59 | _ |
|  | B | 1.02 | 2.60 | 0.00 | 3.62 | -0.62 | _ |
|  | Ga | 0.32 | 1.90 | 9.99 | 12.21 | 0.79 | 2.21 |
|  | Ta | 0.43 | 0.46 | 3.90 | 4.79 | 0.21 | 4.79 |
|  | Ta | 0.43 | 0.46 | 3.90 | 4.79 | 0.21 | 4.79 |
| Ta$_2$GeB$_2$ | B | 1.03 | 2.58 | 0.00 | 3.60 | -0.60 | _ |
|  | B | 1.01 | 2.57 | 0.00 | 3.58 | -0.58 | _ |
|  | Ge | 0.45 | 2.62 | 0.00 | 3.07 | 0.93 | 3.07 |
|  | Ta | 0.43 | 0.47 | 3.98 | 4.87 | 0.13 | 4.87 |
|  | Ta | 0.43 | 0.47 | 3.98 | 4.87 | 0.13 | 4.87 |

### 3.4.4 Theoretical Vickers Hardness

Vickers hardness, derived from the atomic bonds found in solids, indicates how resistant a material is to deformation in extreme circumstances. Several variables influence this feature, such as the crystal flaws, solid structure, atomic arrangement, and bond strength. The Vickers hardness of the M$_2$AB$_2$ (M = Mo, Ta, and A = Ga, Ge) MAX phases is determined using the Mulliken bond population method, as described by Gou et al. [66] using the formula (S7-S10). This method is particularly suitable for partial metallic systems such as MAX phases. Table-5 lists the computed Vickers hardness values for the M$_2$AB$_2$ (M = Mo, Ta, and A = Ga, Ge) compounds. The calculated values are 3.35 GPa, 4.76 GPa, 8.21 GPa, and 8.74 GPa for $Mo_2GaB_2$, $Mo_2GeB_2$, $Ta_2GaB_2$, and $Ta_2GeB_2$, respectively. We observed that $Ta_2GaB_2$ and $Ta_2GeB_2$ have much higher hardness values than $Mo_2GaB_2$ and $Mo_2GeB_2$. These values are also higher than that of other 212 phases, like, $Ti_2InB_2$ (4.05 GPa) [50], $Hf_2InB_2$ (3.94 GPa), and $Hf_2SnB_2$ (4.41 GPa) [26], $Zr_2InB_2$ (2.92 GPa) and $Zr_2TlB_2$ (2.19 GPa) [25], $Zr_2GaB_2$ (2.53GPa), $Zr_2GeB_2$ (3.31GPa), $Hf_2GaB_2$ (4.73GPa) and $Hf_2GeB_2$(4.83GPa) [67]The $H$v calculated by the geometrical average of the individual bonding, where the bonding strength mainly determined by the BOP values. In the case of $Ta_2GaB_2$ and $Ta_2GeB_2$, the BOP of M-B bonding is much higher compared to $Mo_2GaB_2$ and $Mo_2GeB_2$. Even though the BOP of M-B for $Ta_2GaB_2$ and $Ta_2GeB_2$ is higher than that of the other 212 phases mentioned earlier. Therefore, higher values of Hv are expected for $Ta_2GaB_2$ and $Ta_2GeB_2$. Additionally, we looked at the Vickers hardness value between 211 and 212 compounds and discovered that the 212 MAX phase compounds had a higher $H_V$ value. This is because a 2D layer of B atoms is positioned between the M atoms. The B atoms share two center-two electrons to form an extremely strong B-B bond [3].

**Table 5** Calculated data for Mulliken bond number ($n^\mu$), bond length ($d^\mu$), bond overlap populations BOP, ($P^\mu$), metallic populations ($P^{\mu'}$), Vickers hardness ($H_V$) $M_2AB_2$ (M = Mo, Ta, and A = Ga, Ge) compounds.

| Compounds | Bond | $n^\mu$ | $d^\mu$(Å) | $P^\mu$ | $P^{\mu'}$ | $V_b^\mu$ | $H_v$ (Gpa) |
|---|---|---|---|---|---|---|---|
| **$Mo_2GaB_2$** | B1-B2 | 1 | 1.7779 | 2.15 | 0.016 | 5.89 | |
|  | B1-Mo1 | 2 | 2.3217 | 0.17 | 0.016 | 13.12 | 3.35 |
|  | B2-Mo2 | 2 | 2.3217 | 0.16 | 0.016 | 13.12 | |
| **$Mo_2GeB_2$** | B1-B2 | 1 | 1.8059 | 2.15 | 0.031 | 5.96 | |
|  | B1-Mo1 | 2 | 2.3387 | 0.21 | 0.031 | 12.95 | 4.76 |
|  | B2-Mo2 | 2 | 2.3387 | 0.32 | 0.031 | 12.95 | |
|  | Ge1-Mo2 | 2 | 2.6381 | -3.94 | - | - | |
| **$Ta_2GaB_2$** | B1-B2 | 1 | 1.8136 | 2.08 | 0.008 | 6.18 | |
|  | B1-Ta1 | 2 | 2.4113 | 0.59 | 0.008 | 14.53 | 8.21 |
|  | B2-Ta2 | 2 | 2.4113 | 0.54 | 0.008 | 14.53 | |
| **$Ta_2GeB_2$** | B1-B2 | 1 | 1.8488 | 2.08 | 0.008 | 6.22 | 8.74 |
|  | B1-Ta1 | 2 | 2.4345 | 0.63 | 0.008 | 14.21 | |
|  | B2-Ta2 | 2 | 2.4345 | 0.55 | 0.008 | 14.21 | |

|  | | Ge1-Ta2 | 2 | 2.6997 | -0.95 | - | - |  |
|---|---|---|---|---|---|---|---|---|
| [a]**Mo₂GaB** | | B-Mo | 4 | 2.1641 | 1.10 | 0.004 | 27.37 | 3.25 |
| [b]**Zr₂GeB₂** | | B-B | 1 | 1.8456 | 2.3 | 0.020 | 6.510 |  |
|  | | B-Zr | 2 | 2.5072 | 0.19 | 0.020 | 16.31 | 2.53 |
|  | | B-Zr | 2 | 2.5072 | 0.16 | 0.020 | 16.31 |  |
| [c]**Ta₂GaB** | | B-Ta | 4 | 2.2630 | 1.67 | 0.074 | 30.9 | 3.88 |

[a]Reference[16], [b]Reference[45], [c]Reference[35]

## 3.5 Thermal Properties

MAX phases are ideal for high-temperature applications due to their exceptional mechanical qualities at elevated temperatures. As a result, researching the fundamental parameters necessary for predicting their application is of great interest, and these can be obtained from the vibrations of atoms or phonons. The Debye temperature ($\Theta_D$) of a solid is directly connected to its bonding strength, melting temperature, thermal expansion, and conductivity. Using the sound velocity and Anderson's technique [68], the $\Theta_D$ of the phases under study has been computed using the formula (S11). Equation (S12) can be used to get the average sound velocity ($V_m$) from the longitudinal and transverse sound velocities. Equations (S13–14) were used to determine $v_l$ and $v_t$. The calculated values of Debye's temperature are shown in Table 6, where Mo₂GaB₂ has the highest $\Theta_D$ and Ta₂GeB₂ has the lowest. If we rank them, it is as follows: Mo₂GaB₂ < Mo₂GeB₂ < Ta₂GaB₂ < Ta₂GeB₂. Hadi et al. recently reported a MAX phase (V₂SnC) as a TBC material with a $\Theta_D$ value of 472 K [69]. Thus, M₂AB₂ (M = Mo, Ta, and A = Ga, Ge) exhibit encouraging potential as TBC materials, as shown in Table 6.

**Table 6-**Data for density ($\rho$), longitudinal, transverse, and average sound velocities ($v_l$, $v_t$, and $v_m$), Debye temperature ($\Theta_D$), minimum thermal conductivity ($K_{min}$), Grüneisen parameter ($\gamma$), thermal expansion coefficient (TEC) at 300K and melting temperature ($T_m$) of M₂AB₂ (M = Mo, Ta, and A = Ga, Ge) compounds.

| Phases | $\rho$ (kg/m³) | $v_l$ (m/s) | $v_t$ (m/s) | $v_m$ (m/s) | $\Theta_D$ (K) | $K_{min}$ (W/mK) | $\gamma$ | TEC (K⁻¹) | $T_m$ (K) |
|---|---|---|---|---|---|---|---|---|---|
| **Mo₂GaB₂** | 8055 | 6959 | 4032 | 4475 | 585 | 1.20 | 1.49 | 1.52×10⁻⁶ | 2017 |
| **Mo₂GeB₂** | 8227 | 7040 | 3932 | 4378 | 574 | 1.18 | 1.61 | 1.52×10⁻⁶ | 2069 |
| **Ta$_2$GaB$_2$** | 11702 | 5533 | 3185 | 3537 | 448 | 0.88 | 1.51 | 1.47×10$^{-6}$ | 1870 |
| **Ta$_2$GeB$_2$** | 12013 | 5628 | 3112 | 3467 | 438 | 0.89 | 1.65 | 1.48×10$^{-6}$ | 1950 |
| [a]**Ti$_2$PB$_2$** | 4450 | 9480 | 5839 | 6442 | 861 | 1.71 | 1.26 | - | 2033 |
| [b]**Mo$_2$GaB** | 8265 | 6291 | 3498 | 3896 | 480 | 0.940 | 1.63 | - | 1660 |
| [c]**Ta$_2$GaB** | 11890 | 4004 | 999.8 | 1142 | 404.9 | 0.25 | 1.44 | | 1700 |
| [d]**Zr$_2$GaB$_2$** | 6334.81 | 6682 | 4050.28 | 4476.09 | 546 | 0.90 | 1.32 | | 1684 |

[a]Reference[52], [b]Reference[16], [c]Reference[35], [d]Reference [45]

The constant thermal conductivity value at high temperatures is the minimum thermal conductivity ($K_{min}$). As the name implies, this conductivity is minimal because, at high temperatures, phonon coupling breaks. The formula (S15) for the minimum thermal conductivity of solids was derived using the Clarke model [70]. Table 6 lists the calculated value of $k_{min}$, with Ta$_2$GaB$_2$ having the lowest value and Mo$_2$GaB$_2$ having the highest. When selecting suitable materials for TBC applications, a minimum thermal conductivity of 1.25 W/mK is used as a screening criterion [71]. Our compounds exhibit lower minimum thermal conductivity values, holding promising potential as TBC materials. Gd$_2$Zr$_2$O$_7$ and Y$_2$SiO$_5$, two recently developed thermal barrier coating (TBC) materials, have minimal thermal conductivities ($K_{min}$) of 1.22 W/m.K. and 1.3 W/m.K. [72], respectively, as confirmed by experiment. These numbers roughly match the values we computed for M$_2$AB$_2$ (M = Mo, Ta, and A = Ga, Ge).

The Grüneisen parameter ($\gamma$) is a crucial thermal parameter that helps explain the anharmonic effects of lattice dynamics; solids utilized at high temperatures are expected to have lower anharmonic effects. The Grüneisen parameter ($\gamma$) can be determined with the help of Poisson's ratio using equation (S16) [73]. The computed $\gamma$ values, as shown in Table 6, suggest that the compounds under investigation exhibit a weak anharmonic effect. Additionally, for solids with a Poisson's ratio between 0.05 and 0.46, the values similarly fall within the range of 0.85 and 3.53 [74].

The melting temperature ($T_m$) of the compounds under investigation has been calculated using the following formula (S17). The strength of atomic bonding is the primary factor determining the melting temperature of solids; the higher the $T_m$, the stronger the atomic bonding. The order of $T_m$ for the titled phases is found to mirror the $Y$-based (Young's modulus) order, indicating a close link between $T_m$ and $Y$ [75]. As observed in Table 6, our compounds roughly follow the $Y$-based ranking with $Mo_2GeB_2 < Mo_2GaB_2 < Ta_2GeB_2 < Ta_2GaB_2$. The $T_m$ value for $M_2AB_2$ (M = Mo, Ta, and A = Ga, Ge) is also comparable with the TBC material $Y_4Al_2O_9$ (2000 K)[75].

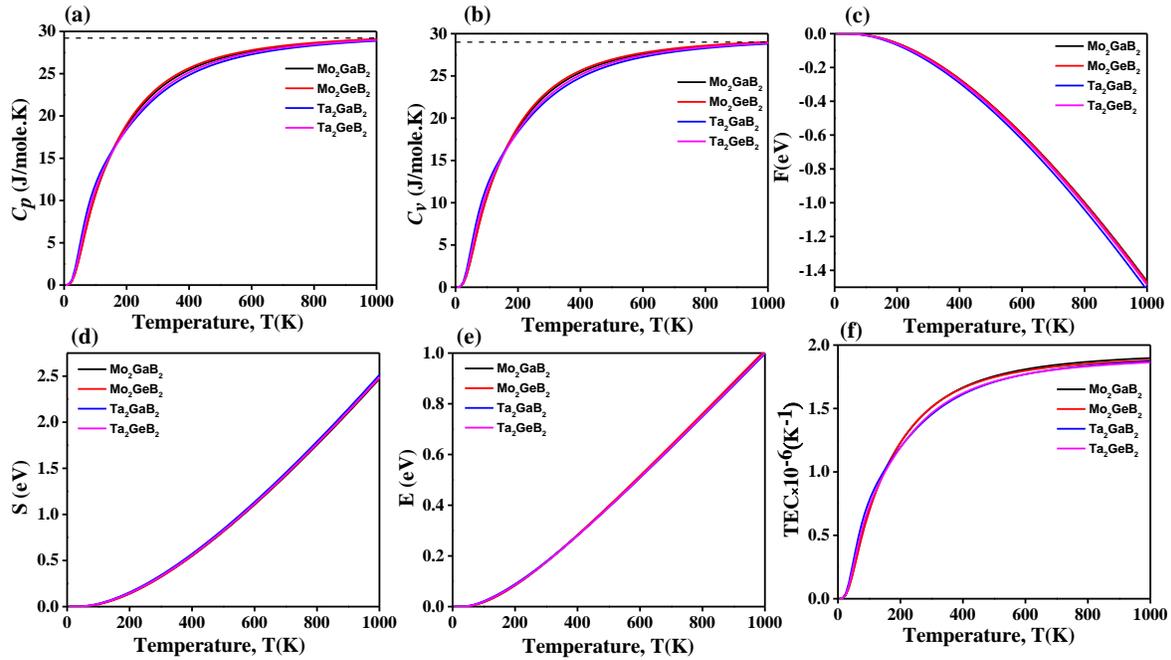

**Fig. 6:** Temperature dependence of the calculated thermodynamic parameters. (a) Specific heat at constant pressure, $C_p$ (b) Specific heat at constant volume, $C_v$ (c) Free energy, $F$ (d) Entropy, $S$ (e) Internal energy, $E$ and (f) Thermal expansion coefficient, $TEC$ of $M_2AB_2$ (M = Mo, Ta, and A = Ga, Ge).

The temperature dependence of specific heats, $C_v$, $C_p$, for the $M_2AB_2$ (M = Mo, Ta, and A = Ga, Ge) compounds was obtained using the formulas (S19-S20), as shown in Fig.6(a,b). Assuming that the quasi-harmonic model is accurate and that phase transitions are not anticipated for the compounds under study, these properties are approximated across a temperature range of 0 to 1000 K. Because phonon thermal softening occurs at higher temperatures, the heat capacity rises with temperature. Heat capacities increase quickly and follow the Debye-$T^3$ power law at lower temperatures. At higher temperature regimes, where $C_v$ and $C_p$ do not greatly depend on temperature, they approach the Dulong-Petit ($3nN_AkB$) limit.[76].

Using the quasi-harmonic approximation, various temperature-dependent thermodynamic potential functions for $M_2AB_2$ (M = Mo, Ta, and A = Ga, Ge) have been estimated at zero pressure and displayed in Fig.6 [77]. These functions include the Helmholtz free energy (F), internal energy (E), and entropy (S) within 0-1000 K temperature using equation (S21-S23). The free energy progressively decreases as the temperature rises, as shown in Fig.6(c). Free energy typically declines, and this trend becomes increasingly negative as a natural process proceeds. As demonstrated in Fig. 6(d), the internal energy (E) shows a rising trend with temperature, in contrast to the free energy. Since thermal agitation creates disorder, a system's entropy rises as temperature rises. This is illustrated in Fig.6(d). A material's thermal expansion coefficient (TEC) is derived from the anharmonicity in the lattice dynamics and can be found using equation (S18). The measure of a material's capacity to expand or contract with heat or cold is called the Thermal Expansion Coefficient, or TEC. As observed in Fig. 5(f), the Thermal Expansion Coefficient (TEC) increases rapidly up to 365 K. Then it approaches a constant value, which indicates lower saturations in the materials with temperature changes. The materials under study have a very low TEC value, a crucial characteristic of materials intended for application in high-temperature technology.

To be effective as thermal barrier coating (TBC) materials, compounds must exhibit a low thermal conductivity ($K_{min}$) to impede heat transfer, a high melting temperature to withstand extreme heat, and a low thermal expansion coefficient (TEC) to maintain dimensional stability under thermal stress. The compounds $M_2AB_2$ (where M = Mo, Ta, and A = Ga, Ge) possess these properties, making them suitable candidates for use as TBC materials

### 3.6 Optical Properties

Different materials exhibit unique behaviors when exposed to electromagnetic radiation. The optical constants determine the overall response of the sample to the incident radiation. The complex dielectric function, defined as $\varepsilon(\omega)=\varepsilon_1(\omega)+i\varepsilon_2(\omega)$, is one of the main optical characteristics of solids. The following formula determines the imaginary part of the dielectric function $\varepsilon_2(\omega)$ from the momentum matrix element between the occupied and unoccupied electronic states.

$$\varepsilon_2(\omega) = \frac{2e^2\pi}{\Omega\varepsilon_0} \sum_{k,v,c} |\psi_k^c|u.r|\psi_k^v|^2 \delta(E_k^c - E_k^v - E)$$

In this formula, *e* stands for an electronic charge, *ω* for light angular frequency, *u* for the polarization vector of the incident electric field, and $\psi_k^c$ and $\psi_k^v$ for the conduction and valence band wave functions, respectively, at k. The Kramers-Kronig equation can estimate the real part of the dielectric function, $\varepsilon_1(\omega)$. In contrast, $\varepsilon_2(\omega)$ and $\varepsilon_1(\omega)$ are utilized to evaluate all other optical parameters, such as the absorption coefficient, photoconductivity, reflectivity, and loss function [78]. In this part, several energy-dependent optical properties of $M_2AB_2$ (M = Mo, Ta, and A = Ga, Ge) 212 MAX phases are calculated and analyzed in detail for the photon energy range of 0 to 30 eV, for [100] plane enabling the first assessment of the compounds' practical applicability.

Given the metallic conductivity of the MAX compounds' electronic structure, additional parameters were chosen to analyze the optical properties. These include a plasma frequency of 3 eV, damping of 0.05 eV, and Gaussian smearing of 0.5 eV [79].

Figure 7(a) shows the real component of the dielectric function, $\varepsilon_1$, which exhibits metallic behavior. In metallic systems, $\varepsilon_1$ has a considerably high negative value in the low-energy range, with the real component reaching negative, which aligns with the band structure finding. Fig. 7(b) depicts the imaginary part of the dielectric function, $\varepsilon_2(\omega)$, representing dielectric losses about frequency. $Mo_2GeB_2$ demonstrates the highest peak in the low-energy region, with all compounds approaching zero from above at around 17 eV. This observation confirms Drude's behavior.

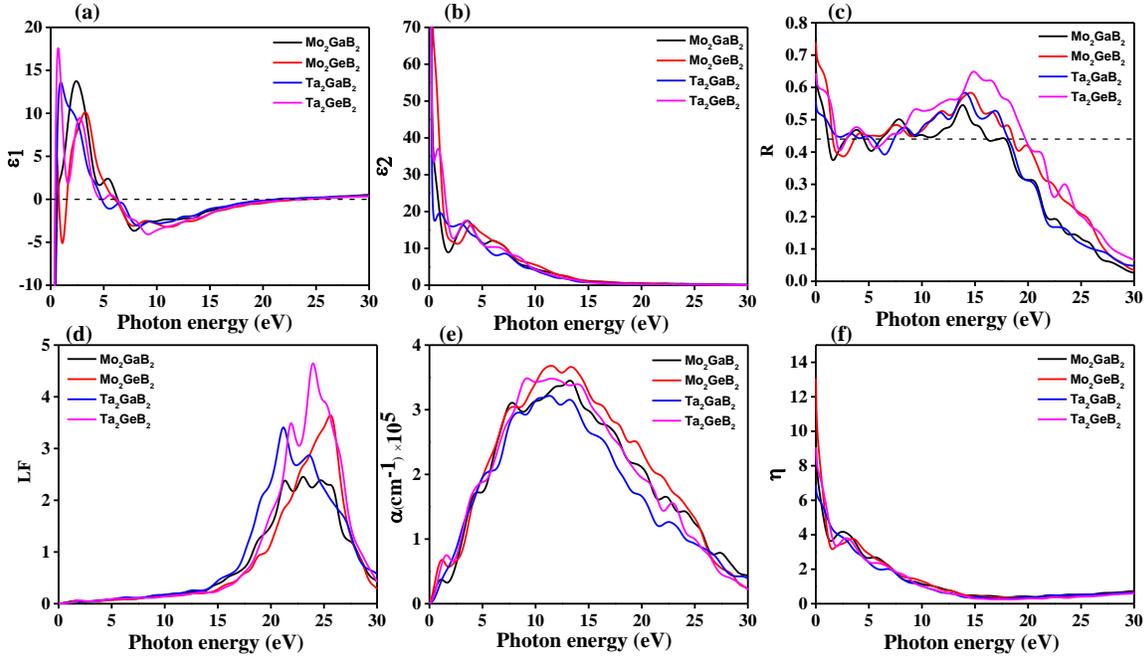

**Fig. 7**- (a) Real part of dielectric constant, $\varepsilon_1$ (b) Imaginary part of dielectric constant, $\varepsilon_2$ (c) Reflectivity, R and (d) Loss function, (e) The coefficient of absorption, ($\alpha$) (f) Refractive index, (*n*) of $M_2AB_2$ (M = Mo, Ta, and A = Ga, Ge) for [100] electric field directions.

The reflectivity of a material measures the percentage of incident light energy reflected off it. Equation (S24) was employed to compute reflectivity using the dielectric function, as shown in Fig.7(c). The spectra's visible and infrared (IR) portions consistently exhibit values exceeding 55%. Consequently, the materials under examination are expected to appear metallic gray. In the infrared (IR) region, the maximum reflectivity for $Mo_2GaB_2$ and $Mo_2GeB_2$ is 65% and 74%, respectively, occurring at 0.33 eV. Conversely, for $Ta_2GaB_2$, the maximum reflectivity is 59% at 14.02 eV, while for $Ta_2GeB_2$, it is 65% at 14.78 eV in the UV region. After 18 eV, the reflectance decreases significantly. According to reports, substances with an average reflectivity value above 44% in the visible light region can effectively reduce solar heating by reflecting a significant amount of the solar spectrum. Among all the compounds we examined, $Ta_2GaB_2$ has a reflectivity in the visible range of more than 44% [78]. Therefore, $Ta_2GaB_2$ can be utilized as a coating material and should be able to mitigate solar heating.

Equation (S25) yielded the energy loss function, LF($\omega$) spectra of the compounds $M_2AB_2$ (M = Mo, Ta, and A = Ga, Ge), which are displayed in Fig.7(d) with the peaks corresponding to plasma frequencies ($\omega_p$). The plasma frequency for $Mo_2GaB_2$, $Mo_2GeB_2$, $Ta_2GaB_2$, and $Ta_2GeB_2$ is observed at 23 eV, 25 eV, 21 eV, and 23 eV, respectively. At this specific frequency, the absorption coefficient rapidly decreases, $\varepsilon_1$ crosses zero from the negative side, and the reflectance $R(\omega)$ displays a falling tail. Above this distinctive frequency, the materials are transparent to the incident electromagnetic radiation.

Fig. 7(e) illustrates the absorbance coefficient ($\alpha$) of compounds $M_2AB_2$ (M = Mo, Ta, and A = Ga, Ge) determined using equation (S26). As $\alpha$ begins to rise from zero photon energy, the metallic nature of the substances under study is again indicated. The visible light region experiences a sharp increase in absorption, peaking in the UV region at around 14 eV and then progressively declining. The IR region exhibits negligible absorption. However, the materials mentioned above appear to have a significant absorption band primarily located in the visible and ultraviolet spectrum. This suggests that the materials can be used in UV surface-disinfection devices, medical sterilization equipment, and other optoelectronic device designs.

Its refractive index is crucial to a material's potential application in optical devices like waveguides and photonic crystals. Equations (S27) and (S28) were used to derive the refractive index ($n$) and extinction coefficient ($k$) for $M_2AB_2$ (M = Mo, Ta, and A = Ga, Ge), as depicted in Fig.7(c) and Fig.S-1(a). The variations of $n$ and $k$ in MAX phase carbides with incident photon energy closely resemble $\varepsilon_1(\omega)$ and $\varepsilon_2(\omega)$. The static refractive index $n(0)$ for $Mo_2GaB_2$, $Mo_2GeB_2$, $Ta_2GaB_2$, and $Ta_2GeB_2$ is 9.4, 13.07, 7 and 9.2, respectively, and decreases gradually with the increase in photon energy. The extinction coefficient for $Mo_2GaB_2$, $Mo_2GeB_2$, $Ta_2GaB_2$, and $Ta_2GeB_2$ gradually increases in the IR region, reaching their maximum values of 2.56, 4.27, 2.42, and 3.28, respectively. Following this, they slowly decrease in the visible and UV regions.

Figure S-1(b) illustrates the photoconductivity ($\sigma$) of $M_2AB_2$ (M = Mo, Ta, and A = Ga, Ge) across various photon energies. The photoconductivity ($\sigma$) parameter quantifies the impact of photon irradiation on a material's electrical conductivity. Similar to the absorbance coefficient ($\alpha$) spectrum, the $\sigma$ spectrum aligns with the metallic nature of $M_2AB_2$ (M = Mo, Ta, and A = Ga, Ge).

## 4 Conclusions

In summary, we employed DFT calculations to explore four 212 MAB phases, $M_2AB_2$ (M = Mo, Ta, and A = Ga, Ge), and investigated the structural, electronic, mechanical, lattice dynamical, and optical properties to predict their possible applications. The phonon dispersion curves, formation energy, and elastic constants collectively suggest that the $M_2AB_2$ boride maintains dynamic, mechanical, chemical, and thermodynamic stability. The electronic band structure and density of states (DOS) offer evidence supporting the metallic nature of the compounds under investigation. Concurrently, the charge density mapping and atomic Mulliken population both confirm the presence of a strong B-B covalent bond. The stiffness constants, elastic moduli, *f*-index, fracture toughness ($K_{IC}$), Pugh's ratio ($G/B$), hardness parameters, and Cauchy Pressure ($CP$) of $M_2AB_2$ were computed and compared with those of their 211 equivalents. We found that the values of the 212-phase borides are higher than those of the 211-phase carbides or borides. $Mo_2GaB_2$ and $Ta_2GaB_2$ are identified as brittle solids, while $Mo_2GeB_2$ and $Ta_2GeB_2$ exhibit ductile characteristics, as indicated by Poisson's ratio (*v*) and Pugh's ratio ($B/G$ or $G/B$). The elastic characteristics display anisotropy due to the distinct atomic configurations along the *a*- and *c*-directions. Vickers hardness calculations are considered reliable indicators of material hardness. The results suggest that $Mo_2GaB_2$ and $Mo_2GeB_2$ possess more pliable characteristics than $Ta_2GaB_2$ and $Ta_2GeB_2$. The high hardness values of $Ta_2GaB_2$ and $Ta_2GeB_2$ compared to the $Mo_2GaB_2$ and $Mo_2GeB_2$ due to higher BOP values of M-B bonding in $Ta_2GaB_2$ and $Ta_2GeB_2$ than in $Mo_2GaB_2$ and $Mo_2GeB_2$. The value of minimum thermal conductivity ($K_{min}$), thermal expansion coefficient, and melting temperature ($T_m$) collectively suggest the potential suitability of $M_2AB_2$ (M = Mo, Ta, and A = Ga, Ge) as a material for thermal barrier coating (TBC) applications in high-temperature devices. The optical conductivity and absorption coefficient corroborate the findings of the electronic band structure. Reflectivity is notably high in infrared (IR) regions and remains nearly constant in the visible and moderate ultraviolet (UV) regions, with an average value exceeding 44% for $Ta_2GaB_2$. This suggests that $Ta_2GaB_2$ can be effectively utilized as a coating material to reduce solar heating. We expect that the comprehensive analysis of the diverse physical characteristics of $M_2AB_2$ (M = Mo, Ta, and A = Ga, Ge) presented in this study will establish a robust foundation for future theoretical and experimental explorations of these fascinating MAB phases.

**CRediT author statement**

A. K. M Naim Ishtiaq and Md Nasir Uddin: Data curation, Writing- Original draft preparation. Md. Rasel Rana, Shariful Islam and Noor Afsary: Reviewing and Editing. Md. Ashraf Ali: Methodology, Reviewing and Editing, conceptualization, supervision; and Karimul Hoque: conceptualization, supervision, editing, and reviewing.

**Declaration of interests**

The authors declare that they have no known competing financial interests or personal relationships that could have appeared to influence the work reported in this paper.

**Acknowledgments**

The authors acknowledge Physics Discipline, Khulna University, Khulna for the logistic support and Advanced Computational Materials Research Laboratory (ACMRL), Department of Physics at Chittagong University of Engineering & Technology (CUET), Chattogram-4349, Bangladesh for laboratory facilities.